# Optimized planning target volume margin in helical tomotherapy for prostate cancer: is there a preferred method?

Yuan Jie Cao, Suk Lee, Kyung Hwan Chang, Jang Bo Shim, Kwang Hyeon Kim, Min Sun Jang, Won Sup Yoon, Dae Sik Yang, Young Je Park, Chul Yong Kim

*Department of Radiation Oncology, College of Medicine, Korea University, Seoul, 136-705*

To compare the dosimetrical differences between plans generated by helical tomotherapy using 2D or 3D margining technique in in prostate cancer. Ten prostate cancer patients were included in this study. For 2D plans, planning target volume (PTV) was created by adding 5 mm (lateral/anterior-posterior) to clinical target volume (CTV). For 3D plans, 5 mm margin was added not only in lateral/anterior-posterior, but also in superior-inferior to CTV. Various dosimetrical indices, including the prescription isodose to target volume (PITV) ratio, conformity index (CI), homogeneity index (HI), target coverage index (TCI), modified dose homogeneity index (MHI), conformation number (CN), critical organ scoring index (COSI), and quality factor (QF) were determined to compare the different treatment plans.

Differences between 2D and 3D PTV indices were not significant except for CI (p = 0.023). 3D margin plans (11195 MUs) resulted in higher (13.0%) monitor units than 2D margin plans (9728 MUs). There were no significant differences in any OARs between the 2D and 3D plans. Overall, the average 2D plan dose was slightly lower than the 3D plan dose. Compared to the 2D plan, the 3D plan increased average treatment time by 1.5 minutes; however, this difference was not statistically significant (p = 0.082). We confirmed that 2D and 3D margin plans are not significantly different with regard to various dosimetric indices such as PITV, CI, and HI for PTV, and OARs with tomotherapy.

E-mail: sukmp@korea.ac.kr

Fax: + 82-2-927-1419

# I. INTRODUCTION

Prostate cancer is the sixth most common cancer in the world, and 85% of cases are diagnosed in men older 65 years [1]. In men with well to moderately-differentiated prostate cancer that remains encapsulated, the clinical progression-free survival rates are 70% and 40% at 5 and 10 years, respectively. In recent years, radiotherapy has been an alternative to radical prostatectomy for localized prostate cancer, as well as the preferred treatment for both locally advanced cases and those in elderly patients [2]. The advent of intensity-modulated radiation therapy (IMRT) has enabled the delivery of highly conformed dose distributions to target volumes in the prostate [3-6]. Helical tomotherapy (HT) is an IMRT delivery technique that provides both highly conformed therapeutic doses as well as image-guidance via megavoltage-computed tomography [7-9]. Dose conformity is critical in radiation therapy since it offers the opportunity to accurately deliver the desired radiation doses to the desired target volume while sparing the nearby healthy organs during treatment.

In practice, however, even HT treatments may be delivered to tissues outside of the prescribed clinical target volume (CTV) due to daily setup uncertainties and changes in patient anatomical structures, especially in cases of prostate [10, 11]. Daily setup uncertainties are typically managed by implementing procedures that reduce setup error; for example, in prostate radiotherapy, a variety of clinical methods is routinely used to manage prostate motion during external beam radiotherapy. These methods include skin marks matched to in-room lasers, weekly or daily ports of the pelvic bony anatomy, trans-abdominal ultrasonography, computed tomography (CT) on rails, intra-prostatic gold seeds visualized with an electronic portal imaging device, and radiofrequency transponders [12-15]. However, since the patient's external position has a limited ability to predict the internal anatomical positions, even with the careful use of these immobilization devices, the target's position may vary from day to day [16, 17]. Furthermore, patients often show notable anatomical changes over the course of treatment due to tumor regression and weight loss, which create relative non-rigidity between individual structures and lead to additional setup error [18, 19].

To account for these setup uncertainties and ensure that a sufficient dose is delivered to the target, treatment planners apply safety margins to the CTV. In clinical applications, both 2 and 3 dimensional (D) automatic margining algorithms can be used to expand the CTV with a constant margin in each of 6 possible directions (left [L], right [R], anterior [A], posterior [P], superior [S], and inferior [I]). The mathematical basis for automatic margining tools is morphological dilation, the equations for which have been discussed elsewhere [20-22]. Methods used by the original 3D automatic margining tools for radiotherapy planning were reported over a decade ago, and their superiority over 2D margining tools and manual planning target volume (PTV) outlining has been discussed [23, 24]. Researchers compared the margining algorithms in commercial treatment planning systems and found significant differences in the 3D margining algorithms of treatment planning systems [25, 26].

However, differences in the automatic 2D and 3D margining methods with regard to planning aspects for HT are not clearly established. The difference in the longitudinal direction could be more significant than in other directions in HT because of the helical treatment method. Here, we compare the dosimetric results of the 2D and 3D automatic margining methods in HT to determine the optimized automatic margining technique for prostate cancer.

## II. Materials and Methods

### A. Patient characteristics

Ten patients previously treated for prostate cancer with HT were selected for this study. The characteristics of these 10 patients are described in Table 1. The patients presented with a variety of stages (stage I to stage III) and target sizes (PTV from 43.7 cm3 to 140.4 cm3). All patients manifested relatively good performance statuses, and none of the treatment areas involved lymph nodes (Table. 1).

### B. CT simulation

The CTV and organs at risk (OARs) were contoured on simulation CT (Philips Medical System, Eindhoven, The Netherlands) images (in-slice spatial resolution, 512 × 512 mm3 and slice interval, 3 mm) for treatment planning. All CT data were transferred to the treatment planning system (TPS, Eclipse version 8.9, Varian Medical System Inc., Palo Alto CA, USA).

### C. Treatment planning

A total of 20 plans, 2 per patient, were generated to compare the differences between 2D and 3D margining algorithms in this study. The planning techniques and plan parameters for the 2D and 3D algorithms were the same for all patients. After contouring all PTVs and OARs in TPS, the planning CT data with radiation therapy structures were transferred from Eclipse TPS to Pinnacle TPS, and the CT data were imported into the TomoTherapy Planning station (Hi-Art version 4.2.1, Accuray Inc., Madison, WI, USA) for tomotherapy IMRT planning.

In the tomotherapy planning system, we added 5-mm margins to CTV in both the 2D and 3D algorithms for each patient, designated PTV-2D and PTV-3D, so that the margin for each PTV was created automatically from both the 2D and 3D margin algorithms, as shown in Fig. 1. The same process was applied to the OARs. In some cases where the OARs overlapped with the PTVs, the OAR volumes were set to the volume that initial OARs substrate the initial PTVs.

All plans were prescribed as 95% PTV receiving 70Gy doses. Tomotherapy plan optimization was performed by a medical physicist according to our protocol. For all plans, the field width, pitch, and modulation factor values were 1.05, 0.287, and 2.5 cm, respectively. Iterations were performed 200 times to satisfy the PTV coverage and OAR tolerance dose levels. The following OARs were defined: bladder, rectum, femoral head, bone marrow, hip joint, and urethra. OAR dose constraints were set according to Emami [27], QUANTEC [28], and RTOG [29]data (Table. 2).

### D. Treatment plan analysis

Several quantitative evaluation tools were used to compare the 20 tomotherapy plans with one another. These included the prescription isodose to target volume (PITV) ratio, homogeneity index (HI), conformity index (CI), target coverage index (TCI), modified dose homogeneity index (MHI), conformity number (CN) for the PTV, maximum dose, mean dose, the dose volume histogram (DVH), and the critical organ scoring index (COSI) for the OAR. PITV ratio was obtained by dividing the prescription isodose surface volume by the target volume [30]. CI, defined as the ratio of the target volume and the volume inside the isodose surface that corresponds with the prescription dose, is generally used to indicate the portion of the prescription dose that is delivered inside the PTV [31]. HI is the ratio of the maximum dose delivered to the PTV and the prescription dose to the PTV [32]. TCI refers to the exact coverage of the PTV in a treatment plan for a given prescription dose. The MHI is similar to the HI, expressed as 95% dose coverage divided by 5% dose coverage [32]. Conformity number (CN) refers to the relative measurement of the dosimetric target coverage and the sparing of normal tissues in a treatment plan [33]. The CN is expressed as:

$$\text{CN} = \text{TCI} \times \text{CI} = \frac{PTV_{PD}}{PTV} \times \frac{PTV_{PD}}{PIV} \quad (1)$$

where $PTV_{PD}$ refers to the PTV coverage at the prescription dose and PIV represents the prescription isodose surface volume. COSI index takes into account both the target coverage and the critical organ irradiation; the main advantage of this index is its ability to distinguish between different critical organs [34]. The COSI is expressed as:

$$\text{COSI} = 1 - \sum_1^n w_i \frac{V_i(OAR)_{>tol}}{TC} \quad (2)$$

where $V(OAR)_{>tol}$ is the fraction of the volume of the OAR that receives more than a pre-defined tolerance dose, and $TC_V$ is the volumetric target coverage, which is defined as the fractional volume of PTV covered by the prescribed isodose. Here, we introduce a dosimetrical index, which may be used to evaluate whole-plan quality, which we refer to as quality factor (QF). The QF of a plan can be analytically expressed as:

$$\text{QF} = [2.718 \exp(-\sum_{i=1}^N W_i X_i)] \quad (3)$$

In the above equation, Xi represents all of the PTV indices used in this study, including PITV, CI, HI, TCI, MHI, CN, and COSI. The values of the weight factor (Wi) can be adjusted between 0 and 1 for all relatively weighted indices for a user-defined number of indices (N). In this study, we used a weighting factor of 1 for all separate indices. Thus, QF was mainly used to compare the conformity of plans in the various trials of a treatment [35].

We used maximum dose, mean dose, and DVHs to quantitatively evaluate the dose distributions in the rectum and bladder. The DVH index employed in our study included $V_5$, $V_{10}$, $V_{20}$, $V_{30}$, $V_{40}$, $V_{50}$, $V_{60}$, and $V_{70}$. A time factor, including the planning time and treatment time, was also included in our study to show treatment and clinical efficacy. In addition, the planning time included the optimization time, and treatment time represented the total treatment time.

### E. Statistical analysis

Data are reported as means ± SDs. To determine whether the differences between dosimetrical and biological indices were significant, the Kruskal-Wallis test and the Mann-Whitney test were performed. All calculations were performed using SPSS software, version 19.0. Differences were considered significant for *p* values < 0.05.

## III. Results

### A. PTV

Isodose distributions of the axial (a, b), coronal (c, d), and sagittal (e, f) for the 2D (left column) and 3D (right column) margin plans are shown in Fig. 2. The orange, cyan, and blue in the PTV indicate color wash areas of 95% (66.5Gy), 50% (35Gy), and, 30% (21Gy), respectively. In a 3-plane comparison (axial, coronal, and sagittal), the 2D and 3D margin plans showed no significant differences in PTV volume isodose distributions. However, when the 3D margin algorithm was applied to the PTV volume, the volume was slightly larger than when the 2D margin algorithm was applied (Fig. 2 c, d). Additionally, DVHs of respective 2D and 3D margin plans were showed in Fig. 3. The results of the dosimetric comparisons of the PTV indices (PITV, CI, HI, TCI, MHI, CN) indicated that most of these indices were better in 3D margining plans than in the 2D plans (Table. 3). Only CI showed a statistically significant difference (p = 0.023). PTV monitor units (MUs) were higher in the 3D margining plan than in the 2D plan (Table. 4), although this difference was not statistically significant (p = 0.076). The 3D margin plan (11195 MUs) increased the MUs by 13.0% compared to 2D margin (9728 MUs). This difference might be due to PTV volume differences between the 2D and 3D margin plans.

### B. OAR

The average OAR doses in the 10 patients are shown in Table. 5, and the DVHs are shown in Fig. 3. The OARs did not differ significantly between the 2D and 3D margin plans. However, the overall average dose in the 2D margin plan was slightly lower than that in the 3D margin plan. This difference was indicated by the COSI values (2D margining plan, 0.9254 ± 0.0603; 2D margining plan, 0.8614 ± 0.1520)

### C. Treatment time

The treatment times of both margining plans are shown in Table. 4. Compared to the 2D margin plan, the 3D margin plan increased the average treatment time by 1.5 minutes; however, this difference was not statistically significant (p = 0.082).

### D. QF

The QF results did not differ significantly between the 2D and 3D plans (Table. 4), although the QF of the 2D plans was higher than that of the 3D plans. (Table 3).

## IV. Discussion

As reported in The International Commission on Radiation Units and Measurements (ICRU) Report 50, the 3D radiation treatment plan technique has become important in radiation therapy [36]. Thus, a study on the appropriate tumor and OAR margins is important with respect to the extent of tumor contouring, the magnitude of patient and organ motion, and treatment setup error [24, 37]. In this paper, therefore, we compared dosimetric results of the 2D and 3D automatic margining HT methods to determine the optimized automatic margining technique for prostate cancer.

The results of this study confirmed that the isodose line of the 3D margin plan gave a larger volume in the axial view than that of the 2D margin plan (Fig. 1 (e)). We showed that the PTV volume for the 3D margin was particularly larger in coronal view (Fig. 2 (c) and (d)). This is because the 2D algorithm plans were created by excluding the superior and inferior direction margins, while the longitudinal direction margin was considered for the 3D margin plan. We confirmed that the 2D margin algorithm had an inherent discrepancy, and the adequate margin algorithm was obtained with the 3D margin method, which did not give a deficient PTV volume. However, the dosimetric results did not differ significantly between the 2 plans according to volume changes in a comparison of dosimetric indices (Table. 3). Although the reason for this is not exactly clear, we thought that this was because in the cases selected were regularly shaped.

The DVHs of the 2 margin algorithm plans appeared to be similar in all OARs (Figure 3). Statistically significant differences were not observed between the 2 methods (Table. 5). However, the 2D margin algorithm plan dose was smaller than that of the 3D margin plan for the low-dose region from V5 to V 20 due to the above-described volume deficiency of the inherent 2D margin algorithm. Our results are consistent with those of the study conducted by Khoo et al. [24], which found that the 2D margin algorithm had a volume deficiency. The deficient volume was an important problem because the PTV and OARs can be underestimated. This could possibly reduce the target coverage and the OAR volume changes. Additionally, this could potentially decrease the likelihood of tumor control and increase the risk of normal tissue complications. Therefore, it is necessary to select an adequate margin algorithm to improve the therapeutic ratio.

Several issues warrant further investigation. In this study, we selected prostate cancer cases with relatively regular tumor shapes. Volume differences between the 2D and 3D margin algorithms due to tumor shape are expected, and thus, we are planning to compare regularly and irregularly shaped tumors. Additionally, further studies are needed to investigate different cases and thus compare the results of this study with those of other cases.

All plans were dosimetrically acceptable (±3%) with regard to target coverage and dose homogeneity, and all critical structures were within the tolerance range. The QF results showed that the combination of smaller field widths and higher pitches might be an important factor for achieving optimized treatment planning parameters. However, it is important to consider time factors for clinical application because smaller field widths can result in longer optimization and treatment times.

## III. CONCLUSION

We confirmed that 2D and 3D margin plans are not significantly different with regard to various dosimetric indices such as PITV, CI, and HI for PTV, and OARs with tomotherapy. To evaluate the optimization plan margin method, we confirmed the feasibility of using QF index. We are planning to investigate the weighted value of each index to improve the accuracy of QF.

## ACKNOWLEDGEMENT

This work was supported by the Korea University Grant.## REREFENCES

1. A. Jemal, A. Thomas, T. Murray, and M. Thun, CA Cancer J Clin, **52**, 23 (2002).

2. G. A. Viani, L. G. da Silva, and E. J. Stefano, Int J Radiat Oncol Biol Phys, **83**, e619 (2012).

3. M. C. Aznar, P. M. Petersen, A. Logadottir, et al., Radiother Oncol, **97**, 480 (2010).

4. I. M. R. T. C. W. Group, Int J Radiat Oncol Biol Phys, **51**, 880 (2001).

5. N. Hardcastle, A. Davies, K. Foo, A. Miller, and P. E. Metcalfe, J Med Imaging Radiat Oncol, **54**,10

Table 1. Characteristics of the patients and tumors.

| Number | | 10 |
|---|---|---|
| Age (years) | <70 | 4 |
| | >70 | 6 |
| T stage | T1 | 2 |
| | T2 | 6 |
| | T3 | 2 |
| N stage | N0 | 10 |
| | N1 | 0 |
| M stage | M0 | 10 |
| | M1 | 0 |
| Grade | I | 2 |
| | II | 6 |
| | III | 2 |
| Gleason score | 6 | 4 |
| | 7 | 1 |
| | 8 | 5 |
| PTV (cc) | | 99.06 |
| Rectum (cc) | | 68.38 |
| Bladder (cc) | | 137.63 |
| Hip joint (cc) | | 30.19 |
| Bone marrow (cc) | | 462.61 |

Table 2. Normal organ tolerance dose.

| Critical structure | QUANTEC* Data | | | | EMAMI¶ Data $TD_{5/5}$ | | |
| --- | --- | --- | --- | --- | --- | --- | --- |
| | Dose(Gy)/Volume(%) | Toxicity rate | Toxicity endpoint | Notes[‡] | Whole | 2/3 | 1/3 |
| Rectum | $V_{50}$ <50% | <10% | Grade 3+ toxicity | Prostate cancer treatment 3D-CRT | 6000 | – | – |
| | $V_{60}$ <35% | | | | | – | – |
| | $V_{65}$ <25% | | | | | – | – |
| | $V_{70}$ <20% | | | | | – | – |
| | $V_{75}$ <15% | | | | | – | – |
| Bladder | $V_{65}$ <50% | – | Grade 3+ toxicity | Prostate cancer treatment (RTOG) 0415 recommendation 3D-CRT | 6500 | 8000 | N/A |
| | $V_{70}$ <35% | – | | | | | – |
| | $V_{75}$ <25% | | | | | | – |
| | $V_{80}$ <15% | – | | | | | – |

Abbreviations: 3D-CRT = 3-dimensional conformal radiotherapy, RTOG = Radiation Therapy Oncology Group.

* All data are estimated from the literature summarized in the QUANTEC reviews unless otherwise noted. Clinically, these data should be applied with caution. Clinicians are strongly advised to use the individual QUANTEC articles to check the applicability of these limits to the clinical situation at hand. They largely do not reflect modern IMRT.

‡ All at standard fractionation (i.e., 1.8–2.0 Gy per daily fraction) unless otherwise noted. Vx is the volume of the organ receiving x Gy. Dmax = Maximum radiation dose.

¶ Emami B, Lyman J, Brown A, et al. Tolerance of normal tissue to therapeutic irradiation. Int J Radiat Oncol Biol Phys 1991;21: 109–122.

Table 3. Index results of 2D and 3D margin plans.

| | 2D | | 3D | | p-value |
|---|---|---|---|---|---|
| | Average | standard deviation | Average | standard deviation | |
| PITV | 0.9976 | 0.0014 | 0.9971 | 0.0013 | 0.290 |
| CI | 0.9593 | 0.0012 | 0.9603 | 0.0016 | 0.023 |
| HI | 1.0468 | 0.0048 | 1.0475 | 0.0049 | 0.450 |
| TCI | 0.9570 | 0.0012 | 0.9575 | 0.0019 | 0.364 |
| MHI | 0.9774 | 0.0028 | 0.9775 | 0.0032 | 0.762 |
| CN | 0.9181 | 0.0019 | 0.9195 | 0.0032 | 0.096 |
| COSI | 0.9254 | 0.0603 | 0.8614 | 0.1520 | 0.450 |
| QF | 1.0316 | 0.0086 | 1.0408 | 0.0226 | 0.364 |
| MU | 9728.7 | 1818.5 | 11194.6 | 1711.3 | 0.076 |
| Actrual MF | 3.7 | 0.3 | 3.7 | 0.3 | 1.000 |
| Tx. Time (m) | 697.0 | 128.3 | 800.6 | 120.7 | 0.082 |

Abbreviations: PITV = prescription isodose surface volume to target volume, CI = conformity index, HI = homogeneity index, TCI = target conformity index, MHI = modified dose homogeneity index, CN = conformity number, COSI = critical organ scoring index, QF = quality factor, AMF = actual modulation factor, Tx = treatment.

Table 4. Dosimetric comparison for OARs between 2D and 3D plan.

|  |  | 2D Average (SD) | 3D Average (SD) | p-value 2D vs. 3D |
|---|---|---|---|---|
| Rectum | $V_5$ | 68.02(14.27) | 75(12.4) | 0.190 |
|  | $V_{10}$ | 60.57(15.34) | 68.8(14.2) | 0.143 |
|  | $V_{15}$ | 57.22(16.21) | 65.6(15) | 0.123 |
|  | $V_{20}$ | 54.15(16.64) | 62.9(15.6) | 0.143 |
|  | $V_{30}$ | 48.66(18.07) | 56.1(17.5) | 0.190 |
|  | $V_{50}$ | 23.68(12.6) | 26.1(13.5) | 0.436 |
|  | $V_{60}$ | 8.34(3.46) | 8.7(3.7) | 0.579 |
|  | $V_{70}$ | 0.21(0.2) | 0.2(0.2) | 0.739 |
|  | $D_{Max}$ | 70.82(0.84) | 70.9(0.7) | 0.796 |
|  | $D_{avg}$ | 27.94(8.76) | 31.3(8.5) | 0.218 |
| Bladder | $V_5$ | 85.57(28.18) | 87.9(25.6) | 0.739 |
|  | $V_{10}$ | 80.14(29.51) | 85.4(28.4) | 0.218 |
|  | $V_{15}$ | 75.97(29.5) | 83.4(29.3) | 0.165 |
|  | $V_{20}$ | 72.75(29.75) | 81.2(30) | 0.190 |
|  | $V_{30}$ | 62.04(29.46) | 71.7(30.4) | 0.315 |
|  | $V_{50}$ | 25(13.67) | 27.8(13.9) | 0.631 |
|  | $V_{60}$ | 11.31(5.86) | 11.8(5.5) | 0.739 |
|  | $V_{70}$ | 0.1(0.16) | 0.1(0.2) | 0.684 |
|  | $D_{Max}$ | 70.75(0.64) | 70.8(0.4) | 0.912 |
|  | $D_{avg}$ | 32.81(14.32) | 37.5(13.5) | 0.353 |
| Hip Joint | $V_5$ | 92.61(12.17) | 96.8(6.2) | 0.579 |
| | $V_{10}$ | 87.65(15.52) | 93.1(11.4) | 0.165 |
| | $V_{15}$ | 80.85(18.8) | 86.4(16.2) | 0.353 |
| | $V_{20}$ | 62.4(21.72) | 67.4(21.6) | 0.436 |
| | $V_{30}$ | 0.286(0.33) | 0.3(0.4) | 0.971 |
| | $D_{Max}$ | 30.51(1.38) | 30.6(1.0) | 0.971 |
| | $D_{avg}$ | 19.75(3.91) | 20.9(3.3) | 0.393 |
| Bone Marrow | $V_5$ | 58.07(18.7) | 62.8(19.5) | 0.631 |
| | $V_{10}$ | 51.62(16.64) | 56.2(17.9) | 0.579 |
| | $V_{15}$ | 32.74(9.06) | 36.3(9.8) | 0.280 |
| | $V_{20}$ | 18.47(6.17) | 20.4(6.6) | 0.436 |
| | $V_{30}$ | 5.62(2.59) | 6(2.7) | 0.529 |
| | $V_{50}$ | 0.78(0.83) | 0.8(0.9) | 0.853 |
| | $V_{60}$ | 0.28(0.5) | 0.3(0.6) | 0.853 |
| | $V_{70}$ | 0.04(0.1) | 0.1(0.2) | 0.971 |
| | $D_{Max}$ | 64.86(6.34) | 65.2(6.2) | 1.000 |
| | $D_{avg}$ | 11.32(2.85) | 12.2(3.0) | 0.529 |

Abbreviations: D = dimensional, SD = standard deviation, $V_5$ = volume receiving at least 5 Gy, $V_{10}$ = volume receiving at least 10 Gy, $V_{15}$ = volume receiving at least 15 Gy, $V_{20}$ = volume receiving at least 20 Gy, $V_{30}$ = volume receiving at least 30 Gy, $V_{50}$ = volume receiving at least 50 Gy, $V_{60}$ = volume receiving at least 60 Gy, $V_{70}$ = volume receiving at least 70 Gy, $D_{max}$ = Maximum point dose, $D_{avg}$ = average dose

Figure Captions.

Fig. 1. The 2D and 3D margin planning target volumes for prostate cancer planned by tomotherapy. a) and b); 2D margin. c) and d); 3D margin. e) and f); 2D and 3D margin.

Fig. 2. The isodose distributions of axial (a, b), coronal (c, d), and sagittal (e, f); from the 2D (left column) and 3D (right column) margin plans for 1 prostate cancer patient. The orange, cyan, and blue in the PTV are representative color wash areas of 66.5 Gy (95%), 35 Gy (50%), and 21 Gy (30%), respectively.

Fig. 3. The DVH data from the 2D and 3D margin plans related with Fig. 2. for 1 prostate cancer patient.

Fig. 1.

|        Axial        |       Sagittal       |
|:-------------------:|:--------------------:|

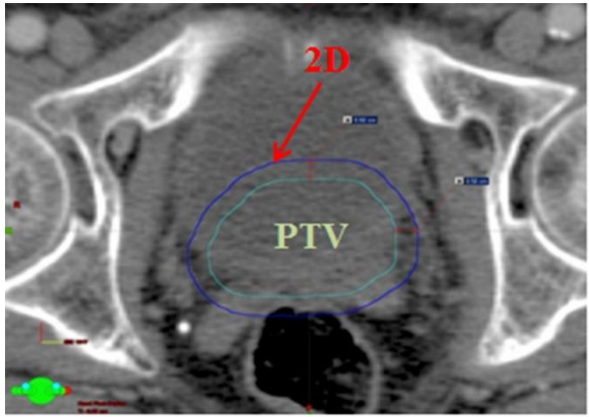
(a)

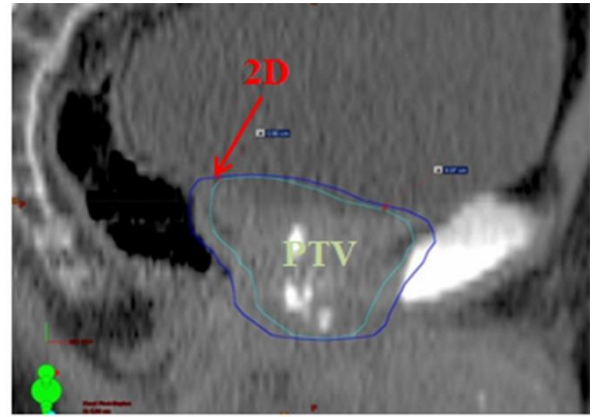
(b)

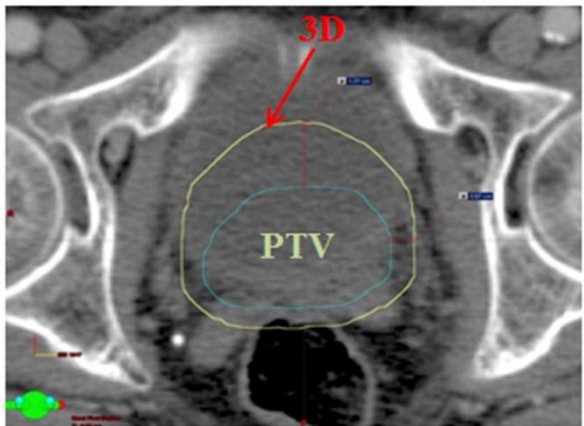
(c)

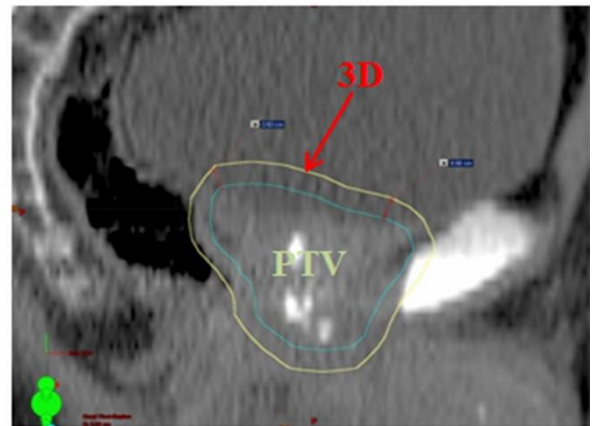
(d)

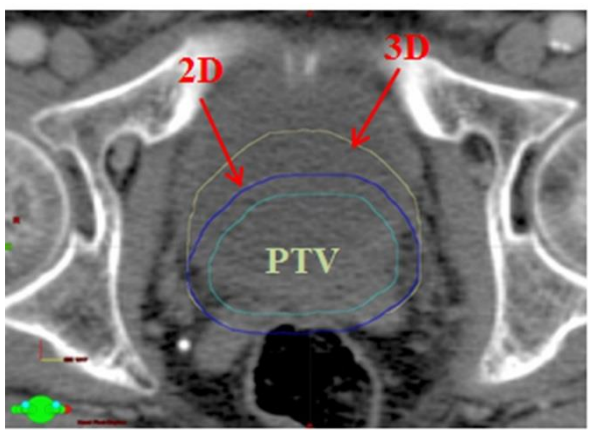
(e)

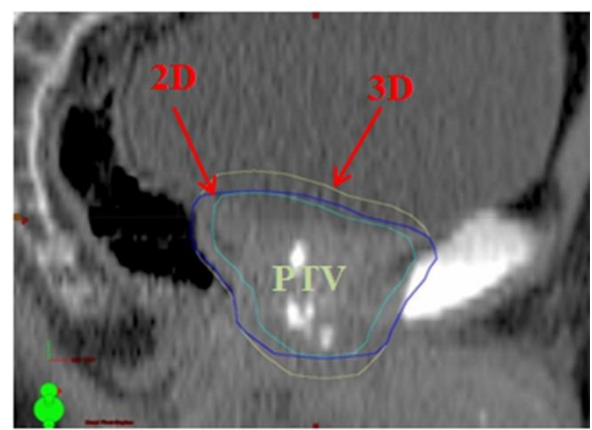
(f)

Fig. 2.

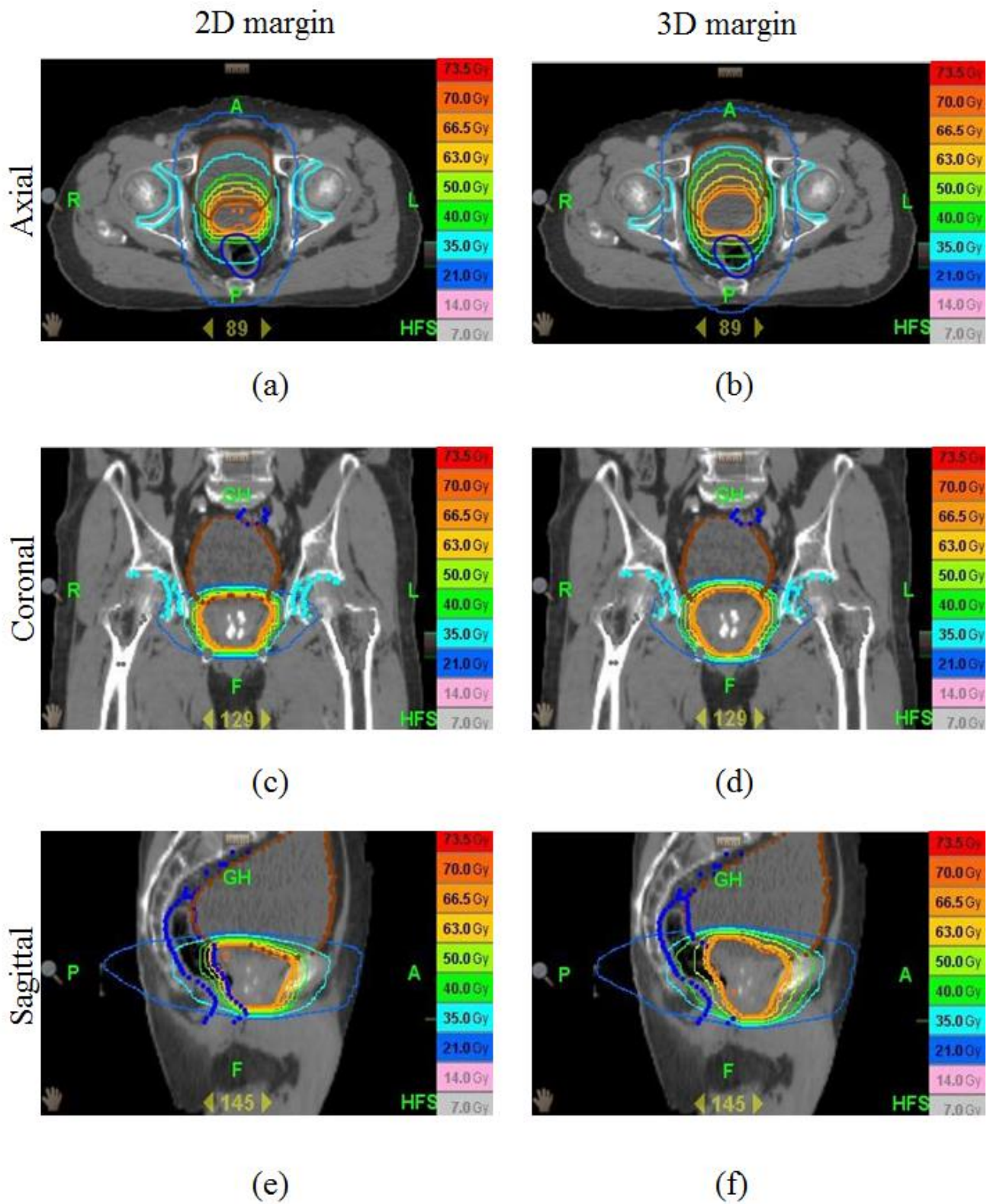

Fig. 3.

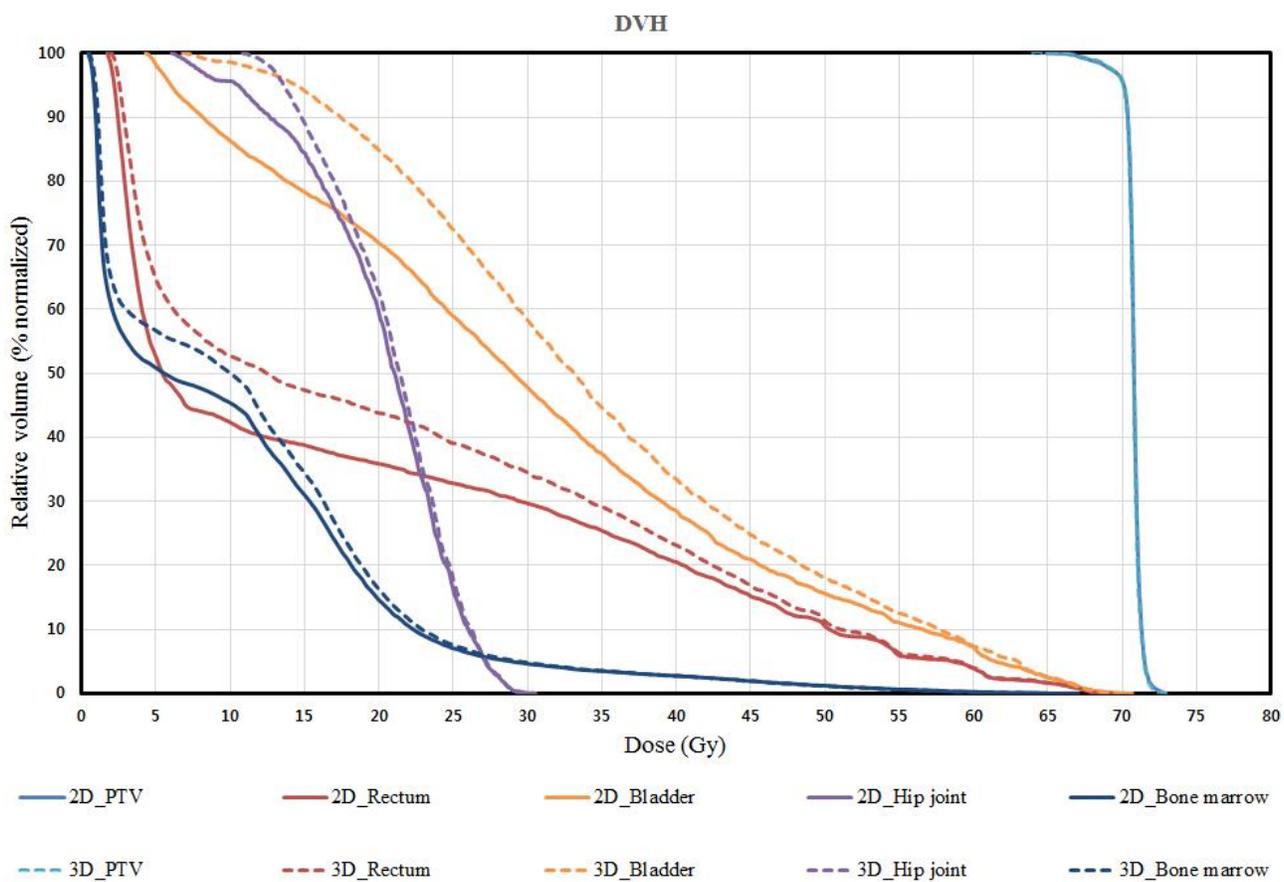